\documentstyle[preprint,aps,tighten]{revtex}

\begin{document}

\title{Dileptons from decay of vector mesons \\
hadronized from a quark-gluon plasma}

\author{Kang Seog Lee$^*$}

\address{Department of Physics, Chonnam National University \\
KwangJu 500-757, S. Korea}

\maketitle

\begin{abstract}
We present a simple model for the dimuon production from the decay of
vector mesons which are hadronized from a baryon-free quark-gluon plasma.
Continuous decay of vector mesons in the medium is known to enhance the dimuon
production and in our model we can estimate the total yield of dimuons decayed
from vector mesons through the rate equations for the number of vector mesons
of interest.
Muon pair production critically depends on the time duration of both the mixed
and hadronic phases. Results are discussed in view of possible
measurements in experiments.

\end{abstract}


\narrowtext

\section{INTRODUCTION}
\label{sec:intro}

Recently dilepton production from vector meson decays in relativistic heavy-ion
collisions has been drawing stong interest\cite{qm93}. One of the reasons is
that the properties of vector mesons such as masses or widths may change in a
hot and dense medium as a result of the onset of chiral symmetry
restoration\cite{hats} and dileptons under vector meson channels may
serve as good probes of such changes\cite{phi,seibert}.

The other reason for the interest is due to the fact that vector mesons
continously decay in the hot and dense medium and thus they may contain
valuable informations on the spacetime evolution of the collision zone.
Heinz and Lee\cite{heinz} suggested to measure the number of dileptons
under rho and omega meson peaks from which one can estimate the lifetime of
the purely hadronic matter. Seibert {\it et al.} suggested to measure the slope
of the transverse mass spectrum of dileptons as a thermometer of the QCD
transition\cite{seibert}.

Or combining the obove two mechanisms there may appear double dilepton peaks
for phi or rho mesons as a consequence of the expected quark-gluon plasma phase
transition. However, in studying the dileptons one relies only on the idea and
still important quantities such as relative magnitude of the two peaks are not
estimable.

In this paper we extend the model by Heinz and Lee\cite{heinz} for the dilepton
production from the purely hadronic matter to the case of the evolution of
quark-gluon plasma and study the total yield of dileptons from the decay of
vector mesons. This model was initailly used to simultaneously explain the
$J/\Psi$ supression and enhancement of $\phi/(\rho+\omega)$ measured by NA35
from a purely hadronic matter and modified to include the free decay to study
the muon pair production\cite{koch}. We believe it is a simple model for the
number
of resonant dileptons which allows easy modification to incorporate the change
of vector meson properties in the medium.

For the dilepton production decayed from vector mesons, $V$, we need to know
the time evolution of the number of vector mesons, $V$, which can be obtained
assuming a model both for the dynamic evolution such as hydrodynamics and for
the hadronization mechanism. Hydrodynamics becomes simple for a baryon-free
quark-gluon plasma whose physical quantities are invariant under longitudinal
Lorentz boosts\cite{bjo} and for the hadronization scenario we assume that
hadronization occurs as a first order phase transition maintaining chemical
equilibrium.  These assumptions lead to a constant rate of hadronization.  By
assuming model equations of state one can evaluate the number of vector mesons
as a function of time starting from an initial conditions reached during the
collisions at a given beam energy.

However, continuous decays of vector meson $V$ may make deviations from
chemical equilibrium and in this paper, we construct rate equations for the
time evolution of vector mesons, $V$, by considering the gain term due creation
of $V$ via secondary collisions and loss terms due to annihilation through
secondary collisions and free decay of $V$ together with the hadronization rate
in the mixed phase. One can solve the rate equations for the particle species
of interest assuming that other particle densities are determined from the
thermal and chemical equilibrium. In this way we loose knowledge on the time
scales of the evolution and will present our results as a function of the time
durations of each phases by controlling the degree of equilibration, namely the
ratio of gain and loss terms.

In section 2, the equilibrium hadronization of a Lorentz-boost invariant
quark-gluon plasma is discussed. Hadronization at constant temperature implies
a constant rate of hadronization during the mixed phase.  In section 3,
considering secondary collisions among hadrons and the free decay of certain
particle species, rate equations for the change in particle numbers are
constructed and solved in both the mixed and hadronic phases.  Muon production
decayed from the particle of interest are formally obtained.  In section 4, the
results of our numerical calculations are presented and we show that dilepton
production from rho meson decays depends critically on the time length of both
the mixed and hadronic phase. We summarize in section 5.

\section{Equilibrium hadronization of a baryon-free quark-gluon plasma}
\label{s2}

In this section we discuss the hadronization rates of particles under the
assumption that thermal and chemical equilibrium is maintained throughout the
phase conversion of a baryon-free quark-gluon plasma.  This assumption
considerably simplifies the calculations, but our final argument does not
qualitatively depend on it. The results depends crucially, however, on our
assumption that the phase transition is of first order. This issue is
still under debate\cite{qm93}.

The most important property of a baryon-free quark-gluon plasma which is
created in the central region during relativistic heavy-ion collisions is the
invariance of the rapidity distribution under longitudinal Lorentz boosts. This
property makes the hydrodynamic description very simple\cite{bjo}:
\begin{equation}
\label{hydro}
 \frac{d^4S}{dx^4} = 0 \qquad \mbox{and} \qquad \frac{d^4N}{dx^4} = 0 .
\end{equation}
Assuming only longitudinal expansion (which initially dominates in any case
[ref]) it is convenient to use the rapidity $y$ and the longitudinal proper
time $\tau$ as varaibles. Then above hydrodynamic equations for the perfect
fluid implies
\begin{equation}
 s(\tau)\tau = s(\tau_0) \tau_0, \qquad \rho(\tau)\tau = \rho(\tau_0) \tau_0 .
\label{inv}
\end{equation}
Here $\rho$ can be the number density of any particle species in the system,
{\it e.g.} the quark density or the baryon density, etc.

In other words, the spatial volume increases linearly in $\tau$ as
a consequence of hydrodynamic expansion:
\begin{equation}
  V(\tau) = \frac{\tau}{\tau_0} V(\tau_0).
\label{vol}
\end{equation}

We assume Eq.~(\ref{inv}) holds throughout the entire evolution of
a quark-gluon plasma, namely through the mixed and hadronic phases until
freeze-out.

As the system cools below due to its hydrodynamic expansion the confinement
phase transition occurs. We assume that this phase transition is of first
order.  Conversion of a quark-gluon plasma to hadronic matter then is usually
described by the mixed phase which is characterized in terms of the volume
fraction $\alpha$, $\alpha=V_H/V_{tot}$.  Here $V_H$ is the volume of the
hadronic subphase in the mixed phase of volume $V_{tot}$.
Then any density, {\it e.g.} entropy density, in the mixed phase can be
 written as
\begin{equation}
 s(\tau) = \alpha(\tau) s_H + (1-\alpha(\tau)) s_Q,
\label{mix}
\end{equation}
where $s_H$ and $s_Q$ are the entropy densities in the hadronic and plasma
subphases, respectively.

If the phase transition maintains chemical equilibrium, entropy and particle
densities can be described via thermal distributions. For a baryon free system,
temperature is the only parameter for the thermal distributions and moreover,
temperature stays constant at $T_c$ during the phase transition. Thus
entropy densities, $s_H$ and $s_Q$, are constant during the phase conversion.
Solving Eq.~(\ref{mix}) for $\alpha(\tau)$ and using Eq.~(\ref{inv}), we get
the time dependence of $\alpha(\tau)$:

\begin{equation}
 \alpha(\tau) = \frac{\tau_H}{\tau_H-\tau_Q} \frac{\tau-\tau_Q}{\tau},
 \label{alpha}
\end{equation}
where $\tau_Q$ and $\tau_H$ are the proper times at the beginnig and end of
hadronization.

Eq.~(\ref{alpha}) implies that as the total volume of the mixed phase
increases linearly in
$\tau$ according to Eq.~(\ref{vol}), the plasma subvolume decreases linearly
while the hadronic subvolume increases linearly.
\begin{eqnarray}
V_Q &=& (1-\alpha(\tau)) V_{tot}(\tau)
    = \frac{\tau_H-\tau}{\tau_H-\tau_Q} V(\tau_Q) \nonumber \\
V_H &=& \alpha(\tau) V_{tot}(\tau)
    = \frac{\tau_H}{\tau_Q} \frac{\tau-\tau_Q}{\tau_H-\tau_Q} V(\tau_Q)
\label{vol1}
\end{eqnarray}

Since the assumption of the chemical equilibrium during the phase conversion
implies that the particle densities are kept constant, the
number of particles hadronized
should also increase linearly in $\tau$. This can be seen as follows:
Since $d^4N/dx^4$
is the particle density with respect to the 4-volume and $dx^4 = \tau d\tau
dyds$, the density of particle species $V$ can be rewritten as
\begin{equation}
 \rho_V(\tau) = \frac{dN_V}{\tau dyds} = \alpha(\tau) \rho_V^H(T_c)
 \qquad \mbox{for} \qquad \tau_Q < \tau < \tau_H .
 \label{dynm}
\end{equation}

One should note that number density in Eq.~(\ref{dynm}), $\rho_V(\tau)$, is
a density with respect to the total volume of the mixed phase,
while $\rho_V^H(T_c)$ is the one with respect only
to the hadronic subvolume. Namely, $ \rho_V(\tau) = \alpha(\tau)
\rho_V^H(T_c)$.
{}From Eq.~(\ref{dynm}) we see that $dN_V/dyds$ indeed grows linearly in
$\tau$.
The corresponding rate equation for the hadronization of particle species $V$
can be written as
\begin{equation}
 \frac{dN_V}{\tau d\tau dyds} = \frac{1}{\tau} \frac{\tau_H}{\tau_H-\tau_Q}
              \rho_V^H(T_c) \equiv \frac{1}{\tau} a_V ,
 \label{hadm}
\end{equation}
with
\begin{equation}
  a_V = \frac{\tau_H}{\tau_H-\tau_Q}\rho_V^H(T_c) ,
\end{equation}
which is constant during the mixed phase since $\rho_V^H(T_c)$ is kept fixed by
its value for the chemical equilibrium at a constant temperature at $T_C$.  For
a baryon-rich quark-gluon plasma, dynamics will be different from
Eq.~(\ref{inv}) and also $\rho_V^H(T_c)$ may vary as $T_C$ changes due to the
reheating.

Eq.~(\ref{hadm}) is a consequence of the gradual phase conversion but, for
different hadronization scenarios, such as suddedn hadronization of strange
particles, the result might be qualitatively different.

In the mixed phase there will also be two-body collisions among the increasing
number of hadrons and free-decays of those hadrons.  In the next section we
consider them explicitly to set up a rate equation and solve it to obtain a
time dependence of the number densities of a vector meson $V$.

\section{Rate equations with secondary collisions and free-decay}
\label{s3}

In a system of many different particle species, collisions between particles
create or annihilate certain particle species $V$. $V$ can decay freely also.
Those processes change the number density of $V$ on top of the hydrodynamic
evolution of the system.
In this section we consider three types of processes which changes the
number of vector meson $V$:
\begin{equation}
 i + j \rightarrow V + X , \qquad \qquad
 l + V \rightarrow X    , \qquad \qquad
 V \rightarrow X  .
\label{coll}
\end{equation}
These processes represent production of vector meson $V$ through a collision
between species $i$ and $j$, annihilation of $V$ via collisions with species
$l$, and the free decay of $V$, respectively.

The rate equation for the number density of particle species $V$ per
unit volume and time can be written as
\begin{equation}
 \frac{dN_V}{d^4x} = \frac{a_V}{\tau}
 + \sum_{i,j} <\sigma v>^{V,X}_{i,j} \rho_i(x) \rho_j(x)
 - \sum_{l} <\sigma v>^{X}_{l,V} \rho_l(x) \rho_V(x)
 - \Gamma_V^{tot} \rho_V(x) ,
\label{rate}
\end{equation}
where the first term on the right hand side is the hadronization rate discussed
in the previous section. The second and third terms are the gain and loss
through the two body collisions and the last term is the loss due to the free
decay as in Eq.~(\ref{coll}).
In Eq.~(\ref{rate}) the scattering cross section $<\sigma v>^{V,X}_{i,j}$
denotes the proper average over the momentum distribution of particles
involved, and the density of particle species $i$, $\rho_i(x)$, depends on time
through its rate equation. Due to the lack of detailed knowledge, however,
we will solve Eq.~(\ref{rate}) only for the vector
mesons of interest and assume that the number densities for other particle
species, which appear in the right hand side of Eq.~(\ref{rate}), are
determined from the chemical equilibrium.

Although Eq.~(\ref{dynm}) looks same in each phases except for the
hadronzation term, its $\tau$-dependence is different for the mixed and
hadronic phase. Thus Eq.~(\ref{rate}) should be solved separately in each
phases, as discussed in the following subsections. It should also be noted that
in the mixed phase $\rho_i(x)$ is the density of particle species $i$ with
respect to the total volume of the mixed phase, as in Eq.~(\ref{dynm}).

\subsection{Mixed phase}
\label{s3a}

In the mixed phase, the particle densities with respect to the total volume can
be written in terms of the densities relative to the hadronic volume as in
Eq.~(\ref{dynm}). Let us define $\lambda_V^{P,M}$ and $\lambda_V^{A,M}$ as
\begin{equation}
\lambda_V^{P,M} = \sum_{i,j} <\sigma v>^{V,X}_{i,j} \rho_i^H \rho_j^H ,\\
\lambda_V^{A,M} = \sum_l <\sigma v>^{X}_{l,V} \rho_l^H ,
 \label{lambdam}
\end{equation}
where superscript $M$ which denotes the "mixed phase" is used to distinguish
similar quantities defined for the hadronic phase.
In Eq.~(\ref{lambdam}) all the densities are taken as those relative to the
hadronic subvolume. For the hadronization maintaing chemical equlibrium,
$\rho_i^H=\rho_i^H(T_c)$ depends only on the transition temperature, and
$\lambda_V^{P,M}$ and $\lambda_V^{A,M}$ depend only on the collision geometry,
but not on the time.

Then, using the relation $d^4x=\tau d\tau dyds$, we can simplfy the rate
equation, Eq.~(\ref{rate}) as
\begin{equation}
 \frac{d}{d\tau} \left(\frac{dN_V}{dyds}\right) = a_V + \tau \alpha^2(\tau)
  \lambda^{P,M}_V - (\alpha(\tau) \lambda^{A,M}_V +
  \Gamma_V^{tot}) \frac{dN_V}{dyds},
\label{ratem1}
\end{equation}

Eq.~(\ref{ratem1}) can be easily solved as
\begin{eqnarray}
 \frac{dN_V}{dyds} (\tau) &=& \tau^{c_{\alpha}\lambda_V^{A,M}\tau_Q}
 \mbox{e}^{-(c_{\alpha}\lambda_V^{A,M}+\Gamma_V^{tot})\tau} \nonumber \\
 & & \cdot \int_{\tau_Q}^{\tau} [a_V+c_{\alpha}^2\lambda_V^{P,M} \frac{(\tau'-
\tau_Q)^2}{\tau'}] \tau'^{-c_{\alpha}\lambda_V^{A,M}\tau_Q}
 \mbox{e}^{(c_{\alpha}\lambda_V^{A,M}+\Gamma_V^{tot})\tau'} d\tau' \nonumber \\
 & & \qquad \qquad \qquad \mbox{for} \qquad \tau_Q < \tau <\tau_H
\label{dnm}
\end{eqnarray}
where $c_{\alpha} = \tau_H/(\tau_H-\tau_Q)$.

In order to get insight for this complicated looking solution and get numerical
values, let us define a ratio $R^0_V$ between the gain and loss terms of $V$
due to secondary collisions and free-decay as
 \begin{equation}
 R^0_V = \frac{\tau \alpha^2(\tau) \lambda_V^{P,M}}
            {(\alpha(\tau)\lambda_V^{A,M} + \Gamma_V^{tot}) \tau \rho_V(\tau)}
     = \frac{\lambda_V^{P,M}}{\lambda_V^{A,M}}
       \frac{\tau_H \frac{\tau-\tau_Q}{\tau_H-\tau_Q}}{\tau \rho_V(\tau)}
       \frac{1}{1+ \frac{\Gamma_V^{tot}}{\alpha\lambda_{V,M}^A}} .
\label{rmr2}
\end{equation}
$R^0_V(\tau)$ is a $\tau$-dependent parameter, which controls the degree of
balance between the gain and loss terms. For $R^0_V(\tau)=1$, we expect that
$dN/dyds$ should linearly increase as a function of $\tau$.
In our numerical estimation for the Eq.~(\ref{dnm}) and thus for the number of
dimuons, we will approximate $R^0_V(\tau)$ as its value at $\tau=\tau_H$.

\begin{equation}
 R^0_V (\tau_H) = \frac{\lambda_V^{P,M}}{\lambda_V^{A,M}}
       \frac{\tau_H-\tau_Q}{\tau_H a_V}
       \frac{1}{1+ \frac{\Gamma_V^{tot}}{\lambda_{V,M}^A}}.
\label{rm}
\end{equation}

This approximation is mainly due to the lack of the
detailed knowledge on the parameters involved, namely $\lambda_V^{P,M}$ and
$\lambda_V^{A,M}$. However, it allows our estimation of the number of vector
mesons $V$ and the corresponding dimuons produced in the mixed phase from the
analysis of the purely hadronic experimental data. This point will be discussed
later.

Writing $R_V=R^0_V (1+ \frac{\Gamma_V^{tot}}{\lambda_V^{A,M}})$, we
eliminate $\lambda_V^{P,M}$ in terms of $R_V$ in Eq.~(\ref{dnm}).
\begin{eqnarray}
 \frac{dN_V}{dyds} (\tau) &=& a_V
\tau^{c_{\alpha}\lambda_V^{A,M}\tau_Q}
 \exp^{-(c_{\alpha}\lambda_V^{A,M}+\Gamma_V^{tot})\tau} \nonumber \\
 & & \cdot \int_{\tau_Q}^{\tau} [1+R_V \lambda_V^{A,M} c_{\alpha}
\frac{(\tau'-\tau_Q)^2}{\tau'}] \tau'^{-c_{\alpha}\lambda_V^{A,M}\tau_Q}
 \exp^{(c_{\alpha}\lambda_V^{A,M}+\Gamma_V^{tot})\tau'} d\tau'
\label{dnm1}
\end{eqnarray}

\subsection{Hadronic phase}
\label{s3b}

Worked out for the mixed phase, time evolution of particle species $V$ in
the hadronic phase can be obtained from Eq.~(\ref{rate}) omitting the term
for the hadronization, {\it i.e.} the first term in the right hand side.
However, one should note that in the hadronic phase particle density is
governed only from the expansion of the system and inversely proportional to
$\tau$ as in Eq.~(\ref{inv}), while in the mixed phase linear increase in
the hadronic subvolume during the phase conversion and constant rate of
hadronization makes the particle density in the hadronic subvolume constant.

{}From the fact that
\begin{equation}
  \frac{dN_V}{dyds} = \tau \rho_V(x) = const.
 \label{dynh}
\end{equation}
in the hadronic phase and with the definition of $\lambda^H$'s for the hadronic
phase as
\begin{eqnarray}
\lambda_V^{P,H} &=& \sum_{i,j} <\sigma v>^{V,X}_{i,j} (dN/dyds)_i
   (dN/dyds)_j \nonumber \\
\lambda_V^{A,H} &=& \sum_l <\sigma v>^{X}_{l,V} (dN/dyds)_l ,
 \label{lambdah}
\end{eqnarray}
we obtain the rate equation for the particle density of $V$ in the hadronic
phase \cite{heinz} as
\begin{equation}
 \frac{d}{d\tau} \left(\frac{dN_V}{dyds}\right) = \frac{1}{\tau}
  \lambda^{P,H}_V - (\frac{1}{\tau} \lambda^{A,H}_V + \Gamma_V^{tot})
  \frac{dN_V}{dyds},
\label{rateh}
\end{equation}

Solution of Eq.~(\ref{rateh}) is
\begin{eqnarray}
 \frac{dN_V(\tau)}{dyds} & = &\frac{dN_V(\tau_H)}{dyds}
  \exp^{[-\Lambda_V(\tau)]} + \lambda_V^{P,H} \exp^{[-\Lambda_V(\tau)]}
       \int_{\tau_H}^{\tau} \frac{d\tau'}{\tau'} \exp^{[\Lambda_V(\tau')]}
\nonumber\\
    & & \qquad \qquad \mbox{for} \qquad\tau_H < \tau <\tau_f ,
\label{dnh}
\end{eqnarray}
where $\Lambda_V$ is defined as
\begin{equation}
 \Lambda_V(\tau) = \lambda_V^{A,H} \ln \frac{\tau}{\tau_H}
                  + \Gamma_V^{tot} (\tau-\tau_H).
\label{lam}
\end{equation}

Similarly for the mixed phase (Eq.~(\ref{rm})), we define in the hadronic phase
the ratio $R^0_V$ between gain and loss terms for particle species $V$ due to
rescattering and free-decay as
\begin{equation}
 R^0_V = \frac{\lambda_V^{P,H}}{(\lambda_V^{A,H} + \tau \Gamma_V^{tot})
      (dN_V/dyds)} .
\label{rh}
\end{equation}
Loss due to free decay of particle $V$ is included in Eq.~(\ref{rh}), which is
omitted in the definition of $R^0_V$ in Ref.~\cite{heinz}.  The free
decay term is not negligible compared to other terms and inclusion of this term
is consistent when we consider the effect of free decay terms.

We will make approximation in the final calculation by taking the value for
$R^0_V$ at $\tau=\tau_H$ in Eq.~(\ref{rh}).  Then from that fact that
$\lambda_V^{A,M} = \lambda_V^{A,H} /\tau_H$ at $\tau=\tau_H$, it is easy to
check that $R^0_V$ defined in Eq.~(\ref{rmr2}) and Eq.~(\ref{rh}) are single
valued at $\tau=\tau_H$.

Defining $R_V=R^0_V(\tau_H) (1+ \frac{\tau_H
\Gamma_V^{tot}}{\lambda_V^{A,H}})$, as in the mixed phase, Eq.~(\ref{dnh})
becomes \begin{equation}
 \frac{dN_V(\tau)}{dyds} = \frac{dN_V(\tau_H)}{dyds} \left[
  \exp^{[-\Lambda_V(\tau)]} + R_V \lambda_V^{A,H}
  \exp^{[-\Lambda_V(\tau)]} \int_{\tau_H}^{\tau} \frac{d\tau'}{\tau'}
   \exp^{[\Lambda_V(\tau')]}  \right] \\
\label{dnh1}
\end{equation}

After breakup of the system, particles decay freely and we have the
exponential decay law
\begin{equation}
\frac{dN_V}{dyds} (\tau) = \frac{dN_V}{dyds} (\tau_f)
                           \exp [-\Gamma_V^{tot} (\tau - \tau_f)]
     \qquad \mbox{for} \qquad \tau_f < \tau .
\label{dnf}
\end{equation}

\subsection{Production of muon pairs }
\label{s3c}

Having the time dependence of the abundance of vector meson $V$ in all
phases we can calculate the dilepton production as
\begin{eqnarray}
 \frac{dN_V^{l^+l^-}}{dyds}
   &=& \Gamma_V^{l^+l^-} \int d\tau \frac{dN_V}{dyds}(\tau)
    \label{muon1} \\
   &=& \Gamma_V^{l^+l^-} \left[ \int_{\tau_Q}^{\tau_H} d\tau
            \frac{dN_V}{dyds}(\tau)
      + \int_{\tau_H}^{\tau_f} d\tau \frac{dN_V}{dyds}(\tau)
      + \frac{1}{\Gamma_V^{tot}}  \frac{dN_V}{dyds}(\tau_f) \right] ,
\label{muon}
\end{eqnarray}
where $\Gamma_V^{l^+l^-}$ is the partial decay width for the the $l^+l^-$
dilepton channel and $\Gamma_V^{tot}$ is the total decay width of $V$.
The first and second terms in Eq.~(\ref{muon}) are production in the mixed and
hadronic phases and the last term is the number of muon pairs decayed after the
freeze-out.
It should be emphasized that in p+p collisions or when thermalized dense
system is not formed during the heavy-ion collisions, we have only the last
term in Eq.~(\ref{muon}) for the production of muon pairs.

Importance of the first two terms can be visualized from an idealized case when
the gain and loss terms balance each other perfectly, namely for
$R^0_V(\tau)=1$.  In this case the number of particle $V$ in the mixed phase is
a linearly increasing function of $\tau$ and that in the hadronic phase is a
constant:
\begin{eqnarray}
 \frac{dN_V}{dyds} (\tau) &=& (\tau-\tau_Q) a_V =\frac{\tau-\tau_Q}{\tau_H-
     \tau_Q} \frac{dN_V}{dyds} (\tau_H)
         \qquad \mbox{for} \quad \tau_Q<\tau<\tau_H \nonumber \\
\frac{dN_V}{dyds} (\tau) &=& \frac{dN_V}{dyds} (\tau_H)
       \qquad\qquad \mbox{for} \quad \tau_H<\tau<\tau_f.
\label{muons}
\end{eqnarray}

Thus we have for the number of muon pairs
\begin{equation}
 \frac{dN_V^{l^+l^-}}{dyds} = B_V \frac{dN_V}{dyds} (\tau_H) \left[
  \Gamma_V^{tot} (\frac{1}{2} \Delta \tau_M + \Delta \tau_H) + 1 \right] ,
\label{ideal}
\end{equation}
where $B_V = \Gamma_V^{l^+l^-} /\Gamma_V^{tot} $ is the branching ratio of $V$
in vacuum for the $l^+l^-$ dilepton channel and
$\Delta \tau_M = (\tau_H-\tau_Q)$ and $\Delta \tau_H = (\tau_f-\tau_H)$
are the time durations of mixed and hadronic phases, respectively.

The first two terms in Eq.~(\ref{ideal}) is the muon production in the mixed
and hadronic phases, respectively, and the last term is those decayed after
freeze-out. For the ideal case of perfect balance between gain and loss terms
the production of muon pairs in the mixed and hadronic phases relative
to those decayed after freeze-out can be simply estimated by comparing the
magnitude of $\Gamma_V^{tot} (\frac{1}{2} \Delta \tau_M + \Delta \tau_H)$ with
1. Please note the factor $1/2$ in the contribution from the mixed phase which
is due to the average of the linearly increasing number of vector mesons $V$.

Having large total decay width, $\Gamma_{\rho}^{tot} = 0.77$ fm$^{-1}$,
$\rho$ mesons produce more dileptons in the medium than after freeze-out
for relatively small lifetimes of a few fermi,
while for other particles, {\it e.g.} for $\omega$ mesons whose total decay
width is $\Gamma_{\omega}^{tot} = 0.043$ fm$^{-1}$ the contributions from the
mixed and hadronic phases becomes significant only when
$(\frac{1}{2} \Delta \tau_M + \Delta \tau_H) \geq 20$ fm.

For a purely hadronic system one can estimate the lifetime of the hot matter,
$\Delta \tau_H$, if the number of vectors mesons $V$ at freeze-out is known and
the number of dileptons from them are measured. Or if we take the ratio of muon
pair production from $\rho$ and $\omega$, for which we expect to be 1 as
in the $p+p$ collisions
due to the small mass difference and the same quark  contents, we can get
valuable information of $\Delta \tau_H$, as suggested in Ref. \cite{heinz}.

Even though the  dependence of the muon production on the lifetime of the mixed
phase is weaker by half than that in the hadronic phase, muon pairs decayed
from the $\rho$ mesons in the mixed phase may be the dominant one when $\Delta
\tau_Q$ is much larger than $\Delta \tau_H$. This could be the case when
the hadronization transition is of first order.  During the first order phase
transition the mixed phase is expected to be long because large
amount of entropy carried by gluons should convert to the entropy of hadrons.
In this case we expect large enhancement of dimuons under the rho peak.

If the mass shift of vector mesons due to restoration of symmetry occurs,
those dimuons produced in the medium will appear at different invariant mass as
suggested in ref. [\cite{seibert,phi}]. However, what we deal in this paper is
the total number of muon pairs regardless of their invariant mass and is not
affected from the possible mass shift.

\section{Results}
\label{s4}

Eq.~(\ref{ideal}) is a result for the perfect balance between the gain and loss
terms, {\it i.e.} $R^0_V(\tau) =1$ for any $\tau$. For the evaluation of
Eqs.~(\ref{dnm1}), (\ref{dnh1}), and (\ref{muons}) in general, values
for the quantities $\lambda_V^{A,M}$, $\lambda_V^{A,H}$ and $R_V(\tau)$ should
be determined together with the time scales, $\tau_Q$, $\tau_H$, and $\tau_f$.

Following the same arguement by Ref.~\cite{koch}, one can estimate
$\lambda_V^{A,H}$ for rho, omega and phi mesons from the data for $J/\Psi$ and
$\phi/(\rho+\omega)$ ratio, both measured in 200 A$\cdot$GeV O+U and S+U
collisions by NA38. Underlying assumption of the arguement is that both the
suppression of $J/\Psi$ and enhancement of $\phi/(\rho+\omega)$ ratio can be
simultaneouly described by the secondary collisions among hadrons through the
rate equations, Eq.~(\ref{dnh1}) for a purely hadronic matter.

When we rewrite $\lambda_V^{A,H}$ as
\begin{equation} \lambda_V^{A,H} = \gamma_V \lambda_{J/\Psi}^{A,H},
\end{equation}
 where
\begin{equation}
\gamma_V = \frac{\sum<\sigma v>_{lV}^{X} (dN_l/dyds) }
              {\sum<\sigma v>_{lJ/\Psi}^{X} (dN_l/dyds) },
\end{equation}
$\gamma_V$ involves only ratios of suitably weighted absorption cross
sections and it does not depend on the details of the nuclear collision
dynamics.
{}From the analysis of NA38 data  we have $\lambda_{J/\Psi}^{A,H} = 0.9$ and
$\gamma_V$ can be approximated as the ratio of cross sections involved,
$\gamma_{\rho} = \gamma_{\omega} \sim 12$. Thus we get $\lambda_{\rho}^{A,H} =
\lambda_{\omega}^{A,H} \sim 11$. One can also estimate $\lambda_V^{A,M}$ from
the relation $\lambda_V^{A,M}  = \lambda_V^{A,H} /\tau_H$, as both quantities
are approximately independent of $\tau$.

As we have defined the quantity $R_V(\tau)$ in the previous section we can
eliminate $\lambda_V^{P,M}$, and $\lambda_V^{P,H}$ in terms of $R_V(\tau)$.
We further approximate $R^0_V(\tau)$, which is a function of $\tau$, with its
value at at $\tau=\tau_H$, {\it i.e.} $R^0_V(\tau) \sim R^0_V(\tau_H)$ and
take it as a parameter which controlls the degree of equilibration. Namely for
$R^0_V >1 $ creation due to secondary collisions wins over the annihilation by
the secondary collisions and free decay, and for $R^0_V <1 $, vice versa.

Time scales involved, $\tau_Q$, $\tau_H$ and $\tau_f$, should be determined
from the hydrodynamic equations together with a model equations of state for
each phase. Instead of making models we rather satisfy ourselves using those
time scales as variables.

It is convenient to define a ratio
\begin{equation}
 Y_V= \frac{dN_V(\tau)/dyds}{dN_V(\tau_H)/dyds}.
\end{equation}

In Fig.~1 we have plotted $Y_V$ for $\rho$ and $\omega$ mesons with arbitrarily
chosen values, $\tau_Q=5$ fm, $\Delta\tau_M=(\tau_H-\tau_Q)=20$ fm,
$\Delta\tau_H=(\tau_f-\tau_H)=35$ fm. Solid lines are the results of the
approximation $R^0_V(\tau) \sim R^0_V(\tau_H) = 1$ while the dotted lines are
for the ideal case, {\it i.e.} Eq.~(\ref{ideal}).
The approximation $R^0_V(\tau) \sim R^0_V(\tau_H) = 1$ underestimates the loss
due to free decay and one should note the importance of this contribution
especially for the rho mesons.

The unknown overall factor $[dN_V(\tau_H)/dyds]$ can be removed by taking a
ratio of muon production from $\rho$ to $\omega$ mesons. In $p+p$ collisions
$\rho$ and $\omega$ mesons are produced in equal numbers, which can be
understood from the small mass difference and same quark contents.
Execept for the medium effect on the number of the two vector mesons, it seems
natural to approximate that $[dN_V(\tau_H)/dyds]$ for $\rho$ and $\omega$ are
same. Then we get from Eq.~({\ref{ideal}) for the perfect balance

\begin{equation}
X = \frac{dN_{\rho}^{l^+l^-}/dyds}{dN_{\omega}^{l^+l^-}/dyds}
 = \frac{B_{\rho} \left[\Gamma_{\rho}^{tot} (\frac{1}{2} \Delta \tau_M +
     \Delta \tau_H) + 1 \right]}
  {B_{\omega} \left[\Gamma_{\omega}^{tot} (\frac{1}{2} \Delta \tau_M + \Delta
  \tau_H) + 1 \right]}
 \label{X}
\end{equation}
with $B_{\rho}/B_{\omega} =0.65$.

In Fig.~2 we show calculated $X$ as a function of the lifetime of the mixed
phase, $\Delta\tau_M=(\tau_H-\tau_Q)$ for two fixed values of
$\Delta\tau_H=(\tau_f-\tau_H)$.

When $\Delta\tau_M /2 + \Delta\tau_H \ll 20$ fm, we can approximate $X$ as
\begin{equation}
 X \sim \frac{B_{\rho}}{B_{\omega}}
   \left[\Gamma_{\rho}^{tot} (\frac{1}{2} \Delta \tau_M + \Delta \tau_H)
    + 1 \right] ,
 \label{X1}
\end{equation}
which is a linear function of $\Delta\tau_M /2 + \Delta\tau_H$.

Taking $\Delta\tau_M =0$ in Eqs.~(\ref{X}) we get the ratio, $X$
for a purely hadronic system.
\begin{equation}
 X = \frac{B_{\rho}\left[\Gamma_{\rho}^{tot} \Delta\tau_H + 1 \right]}
  {B_{\omega}\left[\Gamma_{\omega}^{tot} \Delta\tau_H + 1 \right]}
\sim  0.65 ( \Gamma_{\rho}^{tot} \Delta \tau_H + 1 ) .
 \label{x1}
\end{equation}

According to Eq.~(\ref{x1}) measurement of $X$ larger than 0.65 implies the
formation of a hot and dense medium in thermal equilibrium.
If the hadronization of a quark-gluon plasma
occurs as a first order phase transition, value of $X$ much larger than 0.65 is
expected.

For a purely hadronic system we get from Eq.~(\ref{x1}) $\Delta\tau_H$ by
measuring the ratio of dileptons from $\rho$ to $\omega$ mesons. Thus careful
anaysis of muon production in the $\rho$ and $\omega$ channels may reveal
valuable informations even on the formation of the hadronic fireball as
suggested by Heinz and Lee\cite{heinz}. In Ref.~\cite{heinz} the
dependence of $X$ on the total transverse energy, $E_T$,is investigated.
In this paper, however, we merely note the fact that
as the beam energy or the total transverse energy increases, the
initial temperature of the hadronic fireball increases and so does
$\Delta\tau_H$.

For the beam energy just above that needed to make a deconfinement
transition, a state of mixed phase with small $\Delta\tau_M$ will be formed and
as the beam energy increases, so rapidly does $\Delta\tau_M$ with more or less
saturated $\Delta\tau_H$. The dependence of $(\Delta\tau_M/2+\Delta\tau_H)$ on
the beam energy will be different from that of $\Delta\tau_H$ at lower
energies.

For high enough beam energies a quark-gluon plasma will be made and
$\Delta\tau_M$ and $\Delta\tau_H$ should be functions of the initial
temperature of the QGP. Thus as the beam energy increases there will be two
characteristic changes in the behavior of $X$ at the energies of barely forming
a state of mixed phase and a quark-gluon plasma.

As in the case of the average hadronic $p_T$ spectra as a function of the beam
energy, characteristic changes in slopes of $X$ as a function of the beam
energy can serve as a signature of QGP. Further studies are needed to
investigate how drastic will be the change.

Under the debate is the change of mass of vector mesons in the hot and dense
medium due to the restoration of chiral symmetry, as discussed in the last
section.  If the mass of vector mesons
decreases due to chiral retoration in the hot and dense medium, dileptons
decayed during the mixed phase will peak at smaller invariant mass.

In this case we may have in the invariant mass spectrum a new peak from the
mixed phase at the mass correspoding to the hadronization transition,
$m_V(T_C,\rho_B)$ together with a peak due to the free decay after freeze-out.
Dileptons decayed in the hadronic phase whose temperature and baryon density
change continuously will spread out in the mass region between
$m_V(T_C,\rho_B)$ and $m_V$ in the free space and can hardly be distinguished
from the background.
However, it should be stressed that even when the mass of vector mesons $V$
changes in the hot and dense medium, our results so far are valid as we have
considered the total number of muons pairs decayed from vector mesons $V$ in
each stages of the system but not the invariant mass spectrum.

There may also a change of the decay width of vector mesons in the hot and
dense medium\cite{broaden,kbroad,haglin}. In this case we need to put $\tau$-
dependent $\Gamma_V^{tot}(\tau)$ in every equations and the relation between
$\tau$ and the temperature and baryon density should be determined from the
dynamics of the fireball.

\section{Summary} \label{s5}

We explore interesting aspects of muon pair production in the relativistic
heavy-ion collisions and have shown that careful measurement of muon pairs
under $\rho$ and $\omega$ peaks may reveal valuable informations on the
formation of quark-gluon plasma.

Especially, the ratio of muon pairs from $\rho$ to $\omega$ mesons, $X$,
carries the information on the lifetime of a fireball either in the hadronic or
quark-gluon plasma phase.  Characteristic changes in $X$ as a function of
increasing beam energies should serve as a signature of formation of a quark-
gluon plasma.

Analysis of muon pair production decayed from vector mesons
is made by solving rate equations which is constructed by considering a
hadronization rate, creation and annihilation of particles via secondary
collisions, and free-decays. For the hadronization rate of vector mesons of
interest we consider a baryon-free quark-gluon plasma which is invariant under
Lorentz boost, and assume equilibrium hadronization at a contant temperature.
This implies a constant rate of hadronization during the mixed phase.

However, inside a hot medium broadening of mass due to frequent collisions
will spread the muon peaks under rho mesons and this broadening of rho peak in
the dimuon spectrum may offset the increase of the rho peak to make to
measurement of the increase hard\cite{broaden,kbroad,haglin}.
Studies of the collisional broadening are needed for any realistic estimation
of the enhancement of dimuons under the rho meson peak from the medium.

\acknowledgements

K.~S.~L. would like to acknowledge valuable discussions with U. Heinz.
Especially the concept of collisional broadening is suggested by him.
The work of K.~S.~Lee is supported by KOSEF and BMFT.

\figure{Fig.1:
Number of vector mesons per unit rapidity and transverse area as a
fucntion of $\tau$. Solid lines are the results of the
approximation $R^0_V(\tau) \sim R^0_V(\tau_H) = 1$ while the dotted lines are
for the ideal case, {\it i.e.} Eq.~(\ref{ideal}). }

\figure{Fig.2:
Ratio of muon production from $\rho$ to $\omega$ mesons as a function
of lifetime of the mixed phase. Here time length of hadronic phase is fixed as
5 fm.}

\end{document}